\documentclass[pre, twocolumn]{revtex4-1}
\usepackage{amsmath}
\usepackage{amsfonts}
\usepackage[utf8]{inputenc}

\begin{document}

\title{Hamiltonian formalism and path entropy maximization}

\author{Sergio Davis, Diego Gonz\'alez}
\email{sdavis@gnm.cl, dgonzalez@gnm.cl}

\affiliation{Grupo de Nanomateriales, Departamento de F\'{i}sica, Facultad de Ciencias,
Universidad de Chile, Casilla 653, Santiago, Chile}

\date{\today}

\begin{abstract}
Maximization of the path information entropy is a clear prescription for
constructing models in non-equilibrium statistical mechanics. Here it is shown that, 
following this prescription under the assumption of arbitrary instantaneous constraints on 
position and velocity, a Lagrangian emerges which determines the most probable
trajectory. Deviations from the probability maximum can be consistently
described as slices in time by a Hamiltonian, according to a nonlinear Langevin
equation and its associated Fokker-Planck equation. The connections unveiled 
between the maximization of path entropy and the Langevin/Fokker-Planck equations 
imply that missing information about the phase space coordinate never decreases in time, 
a purely information-theoretical version of the Second Law of Thermodynamics. 
All of these results are independent of any physical assumptions, and thus valid for any generalized
coordinate as a function of time, or any other parameter. This reinforces the view that the
Second Law is a fundamental property of plausible inference.
\end{abstract}



\maketitle

\section{Introduction}

Jaynes' principle of path entropy maximization (also known as the maximum
caliber principle, or MaxCal for short)~\cite{Jaynes1980} is a clear prescription for
the construction of dynamical models, both in nonequilibrium statistical
mechanics~\cite{Dewar2003, Grandy2008} as well as for any dynamical
process~\cite{Haken1986,Presse2013}.

In this work, we derive consequences of the validity of the maximum caliber
principle for the problem of estimating the trajectory of a coordinate, for
instance, the position of a particle as a function of time. This framework will
allow us to describe either classical mechanical systems under uncertainty (e.g. under
the influence of random forces) or stochastic signals such as time series. Position 
and time are familiar concepts to us, but the results of this work apply for
inferences about any quantity $q(s)$ parameterized by a continuous index $s$,
for instance, in geometrical problems under uncertainty.

In the context of classical mechanics it is unavoidable to connect the maximum
caliber formalism, in which the probability of paths is proportional to the
exponential of an action, $P[x()] \propto \exp(A[x()])$, with Feynman's path
integral formalism in quantum mechanics~\cite{Feynman2005}. A related attempt
to bridge path integrals and classical mechanics is found in Ref.
\cite{Gozzi1989}, although in this case the probability distribution is
explicitly constructed to suppress non-classical paths. Connections between
thermodynamics and dynamical systems (although without invoking the idea of
probabilities of paths) is found in Ref.~\cite{Grmela2013} and Ref.~\cite{Grmela2014}.

The aim of this work is twofold: in the first place, it is interesting to
explore to what extent the structure of the dynamical framework of classical
(in the sense of non-quantum) physics is already contained in the simple idea of 
maximization of path entropy. However, the results of this work may also result in 
powerful tools applicable to the continuous maximum caliber formalism.

\section{The Maximum Caliber principle}

Suppose for an unknown coordinate $x$ described as a function of time $t$ we
only know the expectation of a function $f(x, \dot{x}; t)$ (over the
distribution of possible trajectories $x(t)$) as a function of time $F(t)$, that is,

\begin{equation}
\Big<f(x,\dot{x};t)\Big>_I = F(t),
\label{eq_constraint}
\end{equation}
for every instant in the interval $[t_i, t_f]$. According to the maximum caliber
principle, the optimal assigment of probability for the trajectories maximizes  

\begin{equation}
\mathcal{S} = -\int Dx() P[x()|I]\ln \frac{P[x()|I]}{P[x()|I_0]}.
\end{equation}
where $I$ denotes all the given information about the problem, in particular the
constraint in Eq. \ref{eq_constraint}, and $I_0$ is the complete ignorance
state. The path probability $P[x()|I_0]$ is the \emph{a priori} measure needed
for a consistent definition of entropy in continuous systems. Imposing this constraint 
we can write the resulting probability as

\begin{equation}
P[x()|I] = \frac{1}{Z[\lambda()]}P[x()|I_0]\exp\left(-\int_{t_i}^{t_f} dt \lambda(t) f(x(t), \dot{x}(t); t)\right)
\label{eq_prob_caliber}
\end{equation}
where $\lambda$ is a Lagrange multiplier function. In the following we will
assume, for simplicity of the analysis, that the prior measure $P[x()|I_0]$ is
flat. If this is not the case, it can be absorbed as an extra term in the action
$A[x()]$, to be defined below.

If we define a Lagrangian 

\begin{equation}
L(x, \dot{x}; t) = \alpha\lambda(t)f(x, \dot{x}; t),
\label{eq_lagrangian}
\end{equation}
we can rewrite Eq. \ref{eq_prob_caliber} as 

\begin{equation}
P[x()|I] = \frac{1}{Z[\lambda()]}\exp(-\frac{1}{\alpha}A[x()]),
\label{eq_prob_caliber_action}
\end{equation}
where the functional $A$ is the action, defined as

\begin{equation}
A[x()] = \int_{t_i}^{t_f} dt L(x(t), \dot{x}(t); t),
\label{eq_action}
\end{equation}
and the constant $\alpha$, with dimensions of action, is extracted only to
ensure the exponent in Eq. \ref{eq_prob_caliber_action} is adimensional. Eq.
\ref{eq_prob_caliber_action} is analogous to the probability amplitude assigned to a trajectory in the
Feynman path integral formalism~\cite{Feynman2005} (except the exponent here is real-valued), with
$\alpha$ being a constant analogous to Planck's constant. We see that the most probable path is 
automatically prescribed by the principle of minimum action, and thus is a solution 
of the Euler-Lagrange equation

\begin{equation}
\frac{\partial L}{\partial x}=\frac{d}{dt}\frac{\partial L}{\partial \dot{x}}.
\label{eq_euler}
\end{equation}

It is also immediately clear from this formalism that geometric constraints of
the form $g_k(x(t); t)=0$ will only add terms to the Lagrangian, as prescribed
by classical mechanics~\cite{Lanczos1970},

\begin{eqnarray}
L(x, \dot{x}; t) = \alpha\lambda(t)f(x, \dot{x}; t) + \sum_k \mu_k(t)g_k(x(t); t) \\
\nonumber 
= \alpha\lambda(t)\left[f(x, \dot{x}; t)+\sum_k \tilde{\mu}_k(t)g_k(x(t); t)\right],
\end{eqnarray}
where the $\tilde{\mu}_k=\mu_k/(\alpha\lambda)$ are (rescaled) Lagrange multipliers.
Thus, from the point of view of maximum caliber, there is no conceptual distinction between 
the original Lagrangian and the terms from holonomic constraints: the entire Lagrangian always arises from constraints.

A striking example of this loss of distinction occurs when the quantity $F$ which is
known and imposed as a constraint is the instantaneous joint probability
$P(x(t),\dot{x}(t)|I)$. This constraint can be written as 

\begin{equation}
\Big<\delta(x(t)-x)\delta(\dot{x}(t)-\dot{x})\Big>_I = P(x(t), \dot{x}(t)|I),
\end{equation}
i.e., the constraining function $f(x,\dot{x};t)$ is the product of delta
functions. If the probability in the right hand side is known for all $x$ and
$\dot{x}$ in the state space $(x, \dot{x})$ where the probability is non-zero, the Lagrange 
multiplier function will now depend on $(x,\dot{x}, t)$ and the (normalized) action is

\begin{eqnarray}
\frac{A}{\alpha} = \int_{t_i}^{t_f} dt \int dxd\dot{x} \lambda(x, \dot{x};
t)\delta(x(t)-x)\delta(\dot{x}(t)-\dot{x}) \nonumber \\ =
\int_{t_i}^{t_f} dt \lambda(x(t), \dot{x}(t); t),
\end{eqnarray}
i.e. the Lagrange multiplier function itself becomes the
Lagrangian. This means, although in principle the Lagrangian arising from Eq.
\ref{eq_lagrangian} is arbitrary (determined by the choice of constraining function), 
a particular Lagrangian is singled out, the Lagrange multiplier of the probability.

\section{Expectation of functionals}

In order to obtain additional relations for expectation values of arbitrary functionals 
over the distribution given by Eq. \ref{eq_prob_caliber}, we apply the finite difference method~\cite{Gelfand2000}
and map the continuous dynamical problem into a maximum entropy problem with
finite number of degrees of freedom by discretizing time, that is, replacing the 
continuous trajectory $x(t)$ by a vector of $N$ components, $\vec x=(x_1,
\ldots, x_N)$. Eq. \ref{eq_prob_caliber} then reduces to a maximum entropy
solution

\begin{equation}
P(\vec x|I) = \frac{1}{Z}\exp\left(-\frac{1}{\alpha}A(\vec x)\right).
\end{equation}
with the action now replaced by a scalar field $A(\vec x)$, given by

\begin{equation}
A(\vec x) = \Delta t \sum_{j=1}^N L_j(x_j, (x_{j+1}-x_j)/\Delta t).
\end{equation}

There is a connection between expectation values provided by the conjugate
variables theorem~\cite{Davis2012},

\begin{equation}
\Big<\nabla \cdot \vec{v}(\vec x)\Big>_I = \frac{1}{\alpha}\Big<\vec{v}\cdot\nabla A(\vec x)\Big>_I.
\end{equation}

Let us choose $\vec{v}=\hat{e}_k W(\vec x)$ where $\hat{e}_k$ the unit vector in
along the $k$-th coordinate and the trial field $W$ is of the form

\begin{equation}
W(\vec x) = \Delta t \sum_{j=1}^N \omega_j(x_j, (x_{j+1}-x_j)/\Delta t).
\end{equation}

Then,

\begin{equation}
\Big<\frac{\partial}{\partial x_k}W(\vec x)\Big>_I = \frac{1}{\alpha}\Big<W\frac{\partial}{\partial x_k}A(\vec x)\Big>_I.
\end{equation}

Now, for both functionals $W(\vec x)$ and $A(\vec x)$ we have~\cite{Gelfand2000},

\begin{equation}
\frac{\partial}{\Delta t\partial x_k} \rightarrow \frac{\delta}{\delta x(t)},
\end{equation}

and the functional version of the conjugate variables theorem is given by 

\begin{equation}
\Big<\frac{\delta W}{\delta x(t)}\Big>_I = \frac{1}{\alpha}\Big<W[x(t)]\frac{\delta A}{\partial x(t)}\Big>_I,
\label{eq_cvt_func}
\end{equation}
with $W[x()]$ a trial functional with the form

\begin{equation}
W[x()] = \int_{t_i}^{t_f} dt \omega(x(t), \dot{x}(t); t).
\end{equation}

This is the equivalent of the identity (7.30) in Ref.~\cite{Feynman2005} for the quantum mechanical path integrals.

By explicitly replacing the Lagrangian $L$ and the trial function $\omega$, we
can compactly write Eq. \ref{eq_cvt_func} as 

\begin{equation}
\Big<\hat{E}_t\omega\Big>_I = \frac{1}{\alpha}\Big<W[x()]\hat{E}_tL\Big>_I,
\label{eq_cvt_func_E}
\end{equation}
where we have introduced the operator $\hat{E}_t$ as 

\begin{equation}
\hat{E}_t G(x, \dot{x}; t) = \left(\frac{\partial}{\partial x}-\frac{d}{dt}\frac{\partial}{\partial \dot{x}}\right)G(x, \dot{x}; t).
\end{equation}

Choosing $W[x()]=1$ tells us that

\begin{equation}
\Big<\hat{E}_tL\Big>_I=0
\label{eq_EL}
\end{equation}
for all instants $t$, i.e., the Euler-Lagrange equation is valid in expectation 
over the ensemble of trajectories, not only for the most probable one. This suggests 
treating $\hat{E}_t L$ itself as a random variable with zero expectation for the purpose
of constructing a Langevin equation. However, we need to establish the
predicted correlation $\big<\hat{E}_t L\cdot \hat{E}_{t'} L\big>_I$ of the
quantity $\hat{E}_tL$. Now we use $\omega=\hat{E}_{t'}L$ in Eq. \ref{eq_cvt_func_E} to show that 

\begin{equation}
\Big<\hat{E}_t L\cdot \hat{E}_{t'}L\Big>_I =
\Big<\hat{E}_t(\hat{E}_{t'}L)\Big>_I = \frac{1}{\alpha}\Big<\frac{\delta^2 A}{\delta x(t)\delta x(t')}\Big>_I.
\end{equation}
 
As the action has the form in Eq. \ref{eq_action}, its second functional
derivative is 

\begin{eqnarray}
\frac{\delta^2 A}{\delta x(t)\delta x(t')} = 
\frac{\delta}{\delta x(t')}\left[\frac{\partial L}{\partial
x}-\frac{d}{dt}\frac{\partial L}{\partial \dot{x}}\right]\Big|_t\nonumber \\
= \frac{\partial}{\partial x'}\left(\frac{\partial L}{\partial x}-\frac{d}{dt}\frac{\partial L}{\partial \dot{x}}\right)
-\frac{d}{dt'}\frac{\partial}{\partial \dot{x}'} \left(\frac{\partial
L}{\partial x}-\frac{d}{dt}\frac{\partial L}{\partial \dot{x}}\right)
\end{eqnarray}
and this is only non-zero for $t = t'$, thus in all generality we can write it as

\begin{equation}
\Big<\hat{E}_t L\cdot\hat{E}_{t'}L\Big>_I = 2R(t)\delta(t-t')
\label{eq_EL2}
\end{equation}
with $R(t)$ a function of time to be determined. Here we note that, by
taking the limit $t\rightarrow t'$ in Eq. \ref{eq_EL2}, it follows that $R(t)$
cannot be negative. From Eqs. \ref{eq_EL} and \ref{eq_EL2}, we now formally
write the stochastic differential equation (SDE),

\begin{equation}
\hat{E}_tL = \frac{\partial L}{\partial x}-\frac{d}{dt}\frac{\partial L}{\partial \dot{x}} =
\sqrt{2R(t)}\xi(t)
\label{eq_langevin}
\end{equation}
with $\xi(t)$ an unbiased, uncorrelated ``noise'',
\begin{eqnarray}
\Big<\xi(t)\Big>_I = 0 \\
\Big<\xi(t)\xi(t')\Big>_I = \delta(t-t').
\end{eqnarray}

It is crucial to note at this point that $R(t)$ does not necessarily represent a
``random'' physical influence: as the probability interpretation when using the maximum entropy 
and maximum caliber principles is unavoidably Bayesian, $R(t)$ represents
inaccesible or hidden information. Without this uncertainty, the system
follows the trajectory which minimizes the action (Eq. \ref{eq_langevin} reduces
to the Euler-Lagrange equation, Eq. \ref{eq_euler}). 

Eq. \ref{eq_langevin} is, in general, a second-order SDE, but we can obtain a system of two
first-order SDEs by expressing it in canonical form~\cite{Gelfand2000}. For this we define the Hamiltonian 
$\mathcal{H}$ corresponding to the Lagrangian $L$ as the Legendre transformation

\begin{equation}
\mathcal{H}(x,p)=p\dot{x}-L(x,\dot{x}; t) =
\alpha\lambda(t)\left(\dot{x}\frac{\partial f}{\partial \dot{x}}-f\right)\Big|_{x,p},
\end{equation}
together with the momentum $p=\partial L/\partial \dot{x}$. Now we have defined a phase
space $(x,p)$ and by virtue of the Legendre transformation the following relations hold, 
independently of the form of the Lagrangian,

\begin{eqnarray}
\frac{\partial \mathcal{H}}{\partial x} = -\frac{\partial L}{\partial x} \\
\frac{\partial \mathcal{H}}{\partial p} = \dot{x}.
\label{eq_legendre_transf}
\end{eqnarray}

By introducing the Poisson brackets 

\begin{equation}
\{F, G\} = \frac{\partial F}{\partial x}\frac{\partial G}{\partial p}
-\frac{\partial F}{\partial p}\frac{\partial G}{\partial x}
\end{equation}
we see that the required system of first-order SDEs is a nonlinear Langevin
equation in phase space,

\begin{equation}
\dot{\Gamma_i} = \{\Gamma_i, \mathcal{H}\} + R_{ij}(t)\xi_j(t),
\label{eq_langevin_phase}
\end{equation}
where we have used the Einstein summation convention, $\vec \Gamma=(x,p)$ and
the only non-zero component of the uncertainty matrix $R_{ij}$ is $R_{22}(t)=\sqrt{2R(t)}$ with $\xi_2(t)=\xi(t)$.

Eqs. \ref{eq_langevin} and \ref{eq_langevin_phase} give a strong theoretical
basis to the use of the Langevin formalism, independent of physical
considerations. Most importantly, they support the assumption of uncorrelated noise common in the literature
for non-physical Langevin forces~\cite{Yura2014} as an inevitable requirement if the problem
consists only of instantaneous information about position and velocity. Assuming
any other kind of noise will be inconsistent with the information one has by
means of Eq. \ref{eq_constraint}. 

\section{Time evolution of the phase space probability}

We can obtain the time evolution of the probability distribution in phase space, $P(\vec \Gamma(t)|I)=P$, 
by means of the Kramers-Moyal expansion~\cite{Risken1996},

\begin{equation}
\partial_t P = \sum_{n=1}^\infty \frac{(-1)^n}{n!}
\partial^n_{j_1,\ldots,j_n}\left(D^{(n)}_{j_1,\ldots,j_n}(\vec \Gamma; t)P\right),
\end{equation}
where we again used the Einstein summation convention. If we truncate this
expansion at $n=2$, one is left with a Fokker-Planck equation,

\begin{equation}
\partial_t P + \partial_i (D^{(1)}_i P)=\frac{1}{2}\partial^2_{ij}(D^{(2)}_{ij} P),
\end{equation}
with coefficients 

\begin{eqnarray}
D^{(1)}_i = \lim_{\epsilon \rightarrow 0} \frac{1}{\epsilon}\Big<\Delta
\Gamma_i\Big>_{\vec \Gamma} = \{\Gamma_i, \mathcal{H}\}, \\
D^{(2)}_{ij} = \lim_{\epsilon \rightarrow 0} \frac{1}{2\epsilon}\Big<\Delta
\Gamma_i\Delta \Gamma_j\Big>_{\vec \Gamma} = R_{ik}(t)R_{jk}(t),
\end{eqnarray}
where $\Delta \Gamma_i = \Gamma_i(t+\epsilon)-\Gamma_i(t)$. Therefore we can
analyze the time evolution of the phase space probability $P$ by solving 

\begin{equation}
\frac{\partial P}{\partial t} + \{P, \mathcal{H}\} = R(t)\frac{\partial^2 P}{\partial p^2}.
\label{eq_corte}
\end{equation}

This is a Boltzmann transport equation with collision term proportional to the uncertainty $R(t)$. 
Note that, for an arbitrary Hamiltonian, in the case of complete certainty ($R(t)\rightarrow 0$) we recover 
the Liouville equation. It is useful to write Eq. \ref{eq_corte} in logarithmic form, namely, 

\begin{equation}
\frac{\partial}{\partial t}\ln P + \{\ln P, \mathcal{H}\} = R(t)\left[
\frac{\partial^2}{\partial p^2}\ln P + (\frac{\partial}{\partial p}\ln P)^2\right].
\label{eq_corte_log}
\end{equation}

Multiplying by an arbitrary function $G(x,p; t)$ and taking expectation over $P$ we have 

\begin{eqnarray}
\Big<G\frac{\partial}{\partial t}\ln P\Big>_{I,t} + \Big<G\{\ln
P,\mathcal{H}\}\Big>_{I,t} = \\ \nonumber
R(t)\Big<G\left[\frac{\partial^2}{\partial p^2}\ln P + (\frac{\partial}{\partial p}\ln P)^2\right] \Big>_{I,t}
\end{eqnarray}
FIXME \\

Here we apply the general ``fluctuation-dissipation'' relation for an arbitrary
parameter $\alpha$ of a distribution $P$, 

\begin{equation}
\frac{\partial}{\partial \alpha}\Big<G\Big>_{I,\alpha} = \Big<\frac{\partial G}{\partial
\alpha}\Big>_{I,\alpha} + \Big<G\frac{\partial}{\partial \alpha}\ln P\Big>_{I,\alpha}
\label{eq_fd}
\end{equation}
with $\alpha=t$, and the conjugate variables theorem~\cite{Davis2012}, as 

\begin{equation}
\Big<\frac{\partial G}{\partial \Gamma_i}\Big>_I = -\Big<G\frac{\partial}{\partial \Gamma_i}\ln P\Big>_I
\label{eq_cvt_phase}
\end{equation}

After some algebraic manipulations and repeated use of Eq. \ref{eq_fd} and
\ref{eq_cvt_phase} to eliminate $\ln P$, we obtain a classical (non-quantum) analog of the Ehrenfest theorem~\cite{Shankar1994},

\begin{equation}
\frac{d}{dt}\Big<G\Big>_{I,t} = \Big<\frac{\partial G}{\partial t}\Big>_{I,t} +
\Big<\{G,\mathcal{H}\}\Big>_{I,t} + R(t)\Big<\frac{\partial^2 G}{\partial p^2}\Big>_{I,t}.
\label{eq_ehrenfest}
\end{equation}
which gives us the time evolution of an arbitrary phase space function $G$, and
also provides us with a direct way to compute $R(t)$ from time-dependent averages. 

An important remark we address here is the fact that the expectation values over trajectories 
are completely described by the Lagrangian (Eq. \ref{eq_cvt_func_E}), while the
instantaneous properties, representing the system as succesive ``slices'' in time, are in turn 
described by the Hamiltonian (Eq. \ref{eq_ehrenfest}).

\section{Second law of Thermodynamics}

Replacing $G=\ln P$ in Eq. \ref{eq_ehrenfest} and taking into account that, because of Eq. \ref{eq_fd} and
Eq. \ref{eq_cvt_phase}, $\left<\partial_t \ln P\right>_{I,t} = 0$ and
$\left<\{\ln P, \mathcal{H}\}\right>_{I,t} = 0$, we see that 

\begin{equation}
\frac{dS}{dt} = R(t)\Big<(\frac{\partial}{\partial p}\ln P)^2\Big>_{I,t} \geq 0,
\end{equation}
with $S(t)=-\left<\ln P\right>_{I,t}$ the information entropy of the distribution $P$ at time $t$. Therefore, missing
information about the location in phase space never decreases with time, and its
increase is solely due to the uncertainty. In other words, irreversibility is
equivalent to uncertainty over the trajectory followed by the system, whatever
its origin (e.g. sensibility to initial conditions, external driving).

\section{Concluding remarks}

In summary, we have explored the consequences of constructing a maximum caliber dynamical model using 
only instantaneous information about position and velocity. In the first place,
the problem of prediction of the most probable trajectory reduces to standard
classical mechanics, complete with a Lagrangian, a phase space and a
Hamiltonian. For the description of deviations around the most probable
trajectory, the formalism naturally leads to a nonlinear Langevin equation with
uncorrelated noise in phase space, and its corresponding Fokker-Planck equation.
This Fokker-Planck equation is completely determined by the Hamiltonian of the problem.

Given the striking resemblance of Eqs. \ref{eq_prob_caliber_action}, \ref{eq_cvt_func} and \ref{eq_ehrenfest} to real-valued 
versions of the Feynman path integral formulation of quantum mechanics and the Ehrenfest theorem, respectively, 
it remains to be seen if the idea of path entropy maximization could lead to
their quantum mechanical (complex-valued) versions under suitable constraints,
as well as its connection with Nelson's stochastic
mechanics~\cite{Nelson1966,Gonzalez2014}.

\section{Acknowledgments}

The authors gratefully acknowledge funding from FONDECYT grant number 1140514.
DG also acknowledges funding from CONICYT PhD fellowship 21140914.


\end{document}